\documentclass[prl,twocolumn,superscriptaddress]{revtex4}
\usepackage{amsmath}
\usepackage{graphicx}
\usepackage{epsfig}

\begin{document}

\title{Breakdown of the local density approximation
in interacting systems of cold fermions in strongly anisotropic
traps}

\author{Adilet Imambekov}
\affiliation{Department of Physics, Harvard University, Cambridge, Massachusetts 02138, USA}
\author{C. J. Bolech}
\affiliation{Department of Physics, Harvard University, Cambridge, Massachusetts 02138, USA}
\affiliation{Department of Physics, Rice University, Houston, Texas 77005, USA}
\author{Mikhail Lukin}
\author{Eugene Demler}
\affiliation{Department of Physics, Harvard University, Cambridge, Massachusetts 02138, USA}

\date{\today}

\begin{abstract}
We consider spin-polarized mixtures of cold fermionic atoms on
the BEC side of the Feshbach resonance. We demonstrate that a
strongly anisotropic confining potential can give rise to a
double-peak structure in the axial distribution of the density
difference and a polarization-dependent aspect ratio of the
minority species. Both phenomena appear as a result of the
breakdown of the local density approximation for the phase-separated regime. We speculate on the implications of our findings
for the unitary regime.
\end{abstract}

\maketitle

During the past few years, considerable experimental progress has been
achieved in creating systems of strongly interacting ultracold fermions
\cite{coldferm}.  One of the primary motivations for this
research effort is simulating strongly correlated electron systems
such as high-temperature superconductors and quantum magnets. The
number of particles used in experiments with cold atoms is large but
not macroscopic, so they may also exhibit mesoscopic phenomena.  In
this paper, we discuss several important manifestations of such
mesoscopic effects in spin-polarized mixtures of ultracold fermionic
atoms in the vicinity of a Feshbach resonance
\cite{MITunb,Riceunb,theory,Bulgac,Mueller}.  Focusing on the BEC side of the
resonance, we demonstrate that some of the unusual features observed
in experiments by Partridge {\it et al.} \cite{Riceunb}, including a
double-peak structure in the axial distribution of the density
difference and a polarization-dependent aspect ratio of the minority
species \cite{APSReport}, appear as a result of strong anisotropy of a
confining potential. Our starting point is a rigorous proof that
neither of the two phenomena can take place when the confining
potential is parabolic and the local density approximation (LDA) holds
(this has also been noted in Ref.~\cite{Mueller}). A recent
preprint by Zwierlein and Ketterle \cite{comment} argued that the
unharmonicity of the confining potential should have contributed to the double-peak structure observed in Ref.~\cite{Riceunb}.  In this paper, we
show that even for a harmonic potential, but for a strongly
anisotropic trap, such as the one used in experiments of Partridge
{\it et al.}, beyond-LDA corrections give rise to both the double-peak
structure in the axial density difference and a polarization-dependent
aspect ratio.

In this paper, we consider spin-polarized fermion mixtures on the BEC
side of the Feshbach resonance at $T=0.$  At low temperatures and deep into the
BEC limit, all minority-component fermions are paired, forming stable
bosonic molecules. So the system can be thought of as a Bose-Fermi
mixture, where ``bosons'' are tightly bound molecules and ``fermions''
are excess or unpaired fermions of the majority species. Boson-boson
(i.e., molecule-molecule) and boson-fermion (i.e., molecule-fermion)
scattering lengths can be related to the scattering length
between two species of fermions as \cite{termos, PSS,Gurarie}
\begin{align}
a_{bb}&=0.6a,&
a_{bf}&=1.18a\text{.}
\label{scatterings}
\end{align}

\begin{figure}[t]
\psfig{file=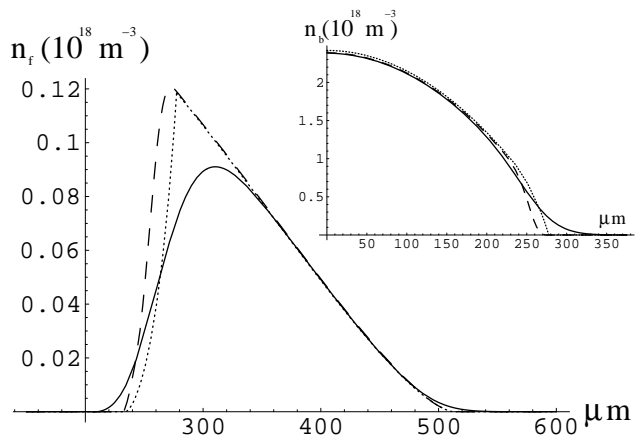} \caption{\label{3ddensities} Rescaled
3D densities of fermions in the Bose-Fermi model
(\ref{initialhamiltonian}).  These correspond to density differences
of the majority and minority components $n_f = n_\uparrow -
n_\downarrow$.  LDA result (dotted), $n_f(r=0,z=x)$ (dashed), and
$n_f(r=x/\Lambda,z=0)$ (solid) for harmonic confinement
$\omega_z/2\pi=7.2\; \mbox{Hz}$ and $\omega_\perp/2\pi=350\; \mbox{Hz}$,
scattering length corresponding to $B=754\;G$,
$N_{\downarrow}+N_{\uparrow}=9.46\times10^4$, and $P=9.5\%.$ Inset:
Bose densities for the same conditions.}
\end{figure}

\begin{figure}[t]
\psfig{file=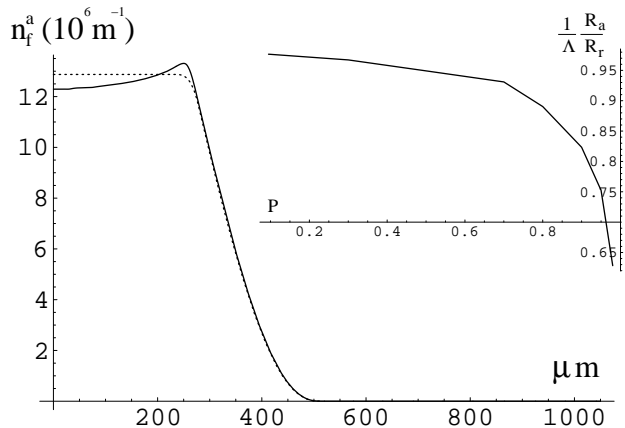} \caption{\label{naxfig} Axial densities of
fermions ($n^a_f = n^a_\uparrow - n^a_\downarrow$) within LDA (dotted)
and in the presence of beyond-LDA corrections (solid), for the same
parameters as in Fig.~\ref{3ddensities}.  The inset shows the
dependence of the anisotropy of the Bose cloud on polarization $P,$
with other parameters being the same as in
Fig.~\ref{3ddensities}. $R_a$ and $R_r$ are obtained from the fits to
the columnar density as discussed in the text. }
\end{figure}

\begin{figure}
\psfig{file=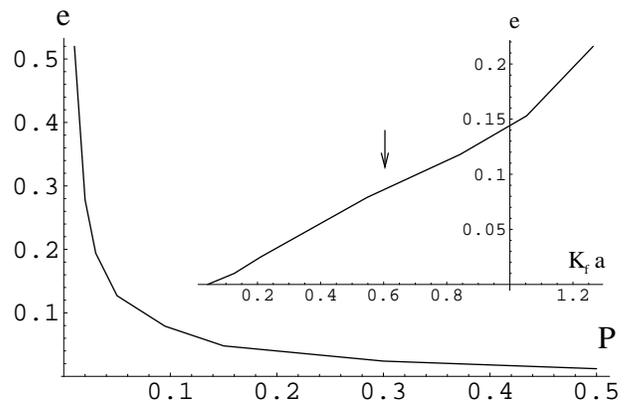} \caption{\label{eversuspa} Dependence of the strength of the double-peak structure $e$  on polarization $P,$ with other parameters being the same as in Fig.~\ref{3ddensities}. The inset shows how $e$ changes with $a$  when we move away from the point considered in Fig.~\ref{3ddensities} (marked with an arrow). $K_f$ is defined as in Ref.~\cite{Riceunb}.}
\end{figure}

Before presenting a formal analysis, we provide a simple qualitative
picture of our results. The leading correction to the LDA comes from
including gradient terms in expressions for the chemical potentials of
bosons and fermions.  The gradient term for bosons corresponds to the
kinetic energy term of the Gross-Pitaevskii equation and has been
thoroughly analyzed in the literature \cite{BECbook}. Gradient terms
for fermions have also been considered in the context of nuclear
physics (see Chap. 4 of Ref.~\cite{Semphys}). Interestingly,
it turns out that gradient terms for fermions have a small numerical
prefactor, and have a negligible effect in the considered region of the
parameters.  For strongly anisotropic traps, gradient terms smooth
the density distribution of bosons in the tightly confined radial
direction at the boundary of the boson cloud (see the inset to
Fig. \ref{3ddensities}). This means that there are ``extra'' bosons in
the radial direction compared to the LDA model. These ``extra'' bosons
interact repulsively with unpaired fermions and push them out in the
axial direction toward the long tips of the trap, leading to a double-peak structure in the axial density of excess fermions (see Fig.
\ref{naxfig}).  We point out that we do not find a critical value of
polarization beyond which the double peak structure appears, although
the peak strength depends on polarization, scattering length, and the
total number of particles in the system (see Fig. \ref{eversuspa}).

The evolution of the aspect ratio of the minority species is also easy
to understand in the BEC limit that we consider. The aspect ratio of
minority species is the aspect ratio of bosons (molecules). When the  LDA
applies, the aspect ratio of bosons is equal to the ratio of confining
frequencies (in experiments reported in Ref. \cite{Riceunb}, this
ratio is around $50$).  In the extreme limit of full polarization, one
can consider a single boson (molecule) in an effective potential
created by the confining potential and excess fermions. Solving a
single-particle Schroedinger equation for the boson gives a
wave function with the aspect ratio equal to the square root of the
ratio of confining frequencies (this corresponds to the extreme
breakdown of the  LDA, see also \cite{gorlitz}). When the number of bosons is small but finite,
one does not find such a dramatic decrease of the aspect ratio, since
even a small number of bosons leads to a change in the fermion
distribution (see the inset to Fig. \ref{naxfig}). However, this argument
explains a general trend of decreasing aspect ratio of minority
species with increasing polarization.

The microscopic model that we consider is the interacting Bose-Fermi
Hamiltonian
\begin{multline}
H=\int d^3r~\Bigl( \frac{\hbar^2}{2(2m)} \nabla \Psi_b^\dagger\nabla \Psi_b
+\frac{\hbar^2}{2m}\nabla \Psi_f^\dagger\nabla \Psi_f \\
+\frac12g_{bb}\Psi_b^\dagger \Psi_b^\dagger\Psi_b\Psi_b
+g_{bf}\Psi_b^\dagger \Psi_f^\dagger\Psi_f\Psi_b  \Bigr)\text{.}
\label{initialhamiltonian}
\end{multline}
Here $\Psi_f\; (\Psi_b)$ is a fermion (boson) operator and $m$ is
the original fermion mass.  The interaction parameters are
$g_{bb}=\frac{2\pi \hbar^2 a_{bb}}{m}$ and
$g_{bf}=\frac{3\pi\hbar^2 a_{bf}}{m}.$ The Hamiltonian
(\ref{initialhamiltonian}) with the parameters
(\ref{scatterings}) can be rigorously derived in the dilute limit
$n_b^{1/3}a\ll1, n_f^{1/3}a\ll1,$ where $n_b$ and $n_f$
are local boson and fermion densities, from a single-channel
model of a wide Feshbach resonance \cite{Gurarie}. It has been
found out numerically \cite{astrakharchik} in the absence of
density imbalance that fermion mixture is well described by the
model of interacting bosons up to $k_f a\lesssim1.$


Within the LDA and assuming a harmonic confinement, the density profiles
are given as solutions of the following equations:
\begin{align}
\mu_f(n_b,n_f)+V(x,y,z)&=\mbox{const}_1\text{,} \label{LDAf}\\
\mu_b(n_b,n_f)+2V(x,y,z)&=\mbox{const}_2\text{,} \label{LDAb}
\end{align}
where $V(x,y,z)=\frac{m \omega_z^2 z^2}{2}+\frac{m \omega_\perp^2
(x^2+y^2)}{2}$. Notice that in the experiment of
Ref.~\cite{Riceunb}, the trap is highly anisotropic
$
\Lambda=\omega_\perp/\omega_z= 48.6 \gg1$.
Boson and fermion densities depend on coordinates ($x$, $y$, $z$)  through the
potential $V(x,y,z)$ only. Hence we should have identical
densities of each species for
points that have the same value of the confining potential.
Thus the aspect ratio of both clouds should always be equal to the
trap anisotropy $\Lambda$.
Then
3D densities along the $z$ axis, $n_b(z)\equiv
n_b(0,0,z)$ and $n_f(z)\equiv n_f(0,0,z)$, provide  complete
information about densities everywhere in the trap.
From 3D densities $n_{b,f}(z)$ one can calculate axial densities as
\begin{multline}
n_f^a(z)=\int dx\,dy\;n_f\left[\sqrt{(x^2+y^2)/\Lambda^2+z^2}\right]\\
=\frac{2\pi}{\Lambda^2}\int_0^\infty r\,dr\; n_f(\sqrt{r^2+z^2})
=\frac{\pi}{\Lambda^2}\int_{z^2}^\infty dt\;n_f(\sqrt{t})\text{.}
\label{naxial}
\end{multline}
For $z>0$, the derivative of the axial density is
\begin{equation}
\frac{dn_f^{a}(z)}{dz}=-\frac{2\pi}{\Lambda^2} z n_f(z)\leq0\text{.}
\label{densder}
\end{equation}
This provides a proof of the absence of peaks in the axial
density away from $z=0$. We note that a similar statement can be made
in the unitary and BCS regimes: if the density of the majority
component is larger than that of the minority component at any point
in the system, then the axial density cannot have a peak except at
$z=0$. Within the LDA and harmonic trapping, phase separation is signaled
not by the appearance of a double-peak structure in the unpaired fermion axial density, but by the existence of an extended region where the axial
density difference has a zero derivative.

We now consider \textit{beyond-LDA} corrections, which arise due
to the spatial derivatives of the densities. At the mean-field
level \cite{Viverit}, correlations between the Bose and Fermi
clouds are neglected and one treats the fermion (boson) density
as providing an external potential for the bosons (fermions).
Including the gradient corrections discussed earlier, up to
second order in gradients of the densities, we find
\cite{BECbook,Semphys}
\begin{multline*}
E=\int d^3r \biggl[ \frac12g_{bb}n_b^2 + g_{bf} n_b n_f + V({\bf
r})(n_f+2n_b)\\ +\frac{\hbar^2}{2m}\Bigl(\frac{|\nabla
\sqrt{n_b}|^2}{2}+\frac{(6\pi^2 n_f)^{5/3}}{10 \pi^2 }+\frac{(\nabla
n_f)^2}{36 n_f}+\frac{\nabla^2 n_f}{3}\Bigr)\biggr]\text{.}
\label{efunctional}
\end{multline*}
In principle, for strong variations of the fermion density, higher-order terms in gradients have to be included. The gradient expansion
above is not suitable for studies of shell structure
\cite{shelleffects}, but since these effects are small in the case we
are studying, this level of approximation is sufficient. Even though
\textit{p}-wave superfluidity of fermions due to boson-induced
interactions has been predicted (see, \textit{e.g.},
Ref.~\cite{Bulgac}), the value of the gap is exponentially small
in $1/n_f^{1/3} a$ and cannot affect significantly the density
profiles obtained from the expression above.

Taking variations of $E$ with the density,
one obtains corrections to the local chemical potential (valid in
the region of nonzero fermion and boson densities),
\begin{align*}
\mu_f&=\frac{\hbar^2}{2m}\biggl[ (6\pi^2n_f)^{2/3}
+\frac{1}{36}\Bigl(\frac{(\nabla n_f)^2}{n_f^2}-\frac{\nabla^2 n_f}{n_f/2}\Bigr)\biggr]
+g_{bf} n_b\text{,}\\
\mu_b&=\frac{\hbar^2}{4m}\frac{\nabla_b^2\sqrt{n_b}}{\sqrt{n_b}} + g_{bb} n_b + g_{bf} n_f\text{.}
\end{align*}
Notice that beyond-LDA corrections to the fermionic chemical potential
carry a small prefactor of $1/36$ compared to the analogous term for
bosons.

To investigate numerically  density profiles in the presence of
kinetic terms, we use the steepest descent method of functional
minimization (used previously for bosons in anisotropic traps
\cite{Dalfovonumerics}). We investigate the effect of beyond-LDA
corrections for typical parameters used in experiments
(cf.~Ref.~\cite{Riceunb}). In Fig.~\ref{3ddensities}, we take
harmonic confinement with $\omega_z/2\pi=7.2\;\mbox{Hz}$ and
$\omega_\perp/2\pi=350\;\mbox{Hz}$, a scattering length that
corresponds to $B=754\;G,$ number of particles
$N=N_{\downarrow}+N_{\uparrow}=2N_b+N_f=9.46\times10^4,$ and
polarization
$P=(N_{\uparrow}-N_{\downarrow})/(N_{\downarrow}+N_{\uparrow})=
N_f/(N_f+2N_b)\sim10\%.$ The LDA density profile is almost completely
phase separated due to the strong repulsion between bosons and
fermions that spatially overlap only in a small region. In
Fig.~\ref{3ddensities}, we compare rescaled 3D density profiles in
radial and axial directions with LDA results, and in Fig.~\ref{naxfig}
we do the same comparison for axial densities and identical values of
the system parameters. Modification of 3D densities can be of the
order of $30\%$ for fermions, and is much stronger in the radial
direction. This is expected, since the gradient corrections in the
radial direction are $\Lambda^2\approx2.4\times10^3$ times larger. We
observe a peak in the axial density, with the density at the maximum
about $10\%$ higher than at $z=0$. The modification of the axial density
is less prominent than changes in the 3D density profile due to the
integration entering the definition of the axial density.
Figure ~\ref{eversuspa} shows the dependence of the relative strength
of the double-peak structure ($e$) on polarization ($P$), where
$e$ is defined as $e=[n^a_f(z)|_{max}-n^a_f(0)]/n^a_f(0)$. It
decreases with increasing $P$, since for a larger polarization a
smaller fraction of the fermion cloud gets affected by the
boundary of the boson cloud. Being essentially a finite-size
effect, $e$ also decreases with increasing total number of
particles. However, this dependence is pretty weak, since the
size of the cloud depends weakly  on the total number of atoms
(only as $N^{1/6}$ for the unitary regime within the LDA). The inset
to Fig.~\ref{eversuspa} shows the dependence of $e$ on scattering
length ($a$). As $a$ increases, the region of spatial overlap of
the Bose and Fermi clouds shrinks due to stronger repulsion, and
this leads to larger gradients and larger beyond-LDA corrections
for the boson cloud. In addition, due to the enhanced repulsion,
the fermions are more sensitive to the corrections to the boson
cloud. Both of  these effects lead to an increase of $e$ for larger
$a.$ We present results as a function of $K_fa,$ where $K_f$ is
defined as in Ref.~\cite{Riceunb}. Notice, however, that
there is no ``universality'' in this curve: indeed, increasing
the total number for fixed $a$ would increase $K_fa$, while $e$
would decrease, as discussed earlier. Although $K_fa$ in
Fig.~\ref{eversuspa} is of the order of $1,$ the main physics we
are interested in takes place at the boundary of the boson cloud,
where local $k_f a$ is considerably smaller, so the treatment of
the system as a Bose-Fermi mixture is well justified.

Beyond-LDA corrections increase the size of the boson cloud in the
radial direction compared to the LDA result, thus reducing the aspect
ratio of the bosonic cloud compared to $\Lambda.$ As the polarization
($P$) increases, for a fixed total number of atoms, the size of the
bosonic cloud decreases and the importance of beyond-LDA effects
grows. The inset to Fig.~\ref{naxfig} shows the evolution of
anisotropy as a function of polarization for harmonic
confinement. $R_a$ and $R_r$ are defined as follows: (i) the columnar
density $n^{c}_b(r,z)$ is obtained from $n_b(r,z)$, assuming radial
symmetry in the $x$-$y$ plane; (ii) one then fits
noninteracting-fermion density distributions to the columnar density
along axial and radial directions,
$n^{c}_b(r,0)=A(1-\frac{r^2}{R_r^2})^2$ and
$n^{c}_b(0,z)=B(1-\frac{z^2}{R_a^2})^2$. Notice these functional forms
are not always suitable and the fits provide just an estimate
of the radii. The aspect ratio of the fermionic cloud does not get
significantly modified in a harmonic confinement.

So far we discussed only the BEC regime, where controllable analytic
theory is available. We developed a consistent qualitative picture of
the appearance of peaks in the axial density difference,
$n^a_f=n^a_\uparrow - n^a_\downarrow$, and a decrease of the aspect
ratio of minority species with increasing polarization.
Quantitatively, the numbers we obtain in the BEC regime are smaller
than what is observed experimentally in Ref.~\cite{Riceunb} for
the unitary regime, but of the same order of magnitude as the effects
of unharmonicity discussed in Ref.~\cite{comment}.  Now we will
comment on the relevance of our findings for the unitary limit,
where most experiments are performed.  Results from the BEC regime
should not be  directly extrapolated toward the unitary regime, since
the treatment of the system as a Bose-Fermi mixture starts to break down, and the form
of beyond-LDA corrections in the unitary regime is not known. Hence
the statements made in this paragraph are only speculative.  One can
estimate, within the LDA, the chemical potential in the center of the trap
based on Fig. 2C of Ref.~\cite{Riceunb}: for a typical
configuration it is $\sim 10\hbar\omega_\perp$, where
$\hbar\omega_\perp$ is the radial level spacing. The chemical
potential at the boundary of the inner cloud is about three to four times
$\hbar\omega_\perp$, whereas the gradient contribution in the radial
direction, which is neglected in the LDA, is
$\sim\hbar^2\Lambda^2/[m(\Delta R)^2]\sim\hbar\omega_\perp$ (where
$\Delta R$ is the difference in axial radii). It is thus clear that
beyond-LDA corrections might also be relevant for the unitary
regime. The appearance of double peaks has been interpreted in
Ref.~\cite{Riceunb} as  evidence for phase separation of
excess fermions from a paired central core. Although peaks should not
appear in the LDA, the picture we have in the BEC limit provides some
support to this interpretation. When excess fermions spatially overlap with the
paired core, they are less sensitive to beyond-LDA corrections, which
are important only at the boundary of the paired cloud. As phase
separation takes place, excess fermions are expelled to the boundary,
where beyond-LDA effects become more important. These corrections are
stronger in the radial direction, hence fermions are pushed in the
axial direction to larger $|z|$. This may give rise to a pronounced
double-peak structure in the axial density. The high anisotropy of the
trap is important for this scenario.

To summarize, we considered unbalanced fermions in the BEC limit for
$T=0.$ We proved a ``theorem'' that prohibits peaks in the axial
density for $z\neq0$ within the LDA and for the harmonic trapping
potential. We showed that for strongly anisotropic confinement,
beyond-LDA corrections produce a double-peak structure in the axial
density, and change the aspect ratio of the inner cloud. We discussed
the implications of our findings for the unitary regime.

We thank R.~Hulet, M.~Zwierlein, W.~Ketterle, and H.~Stoof for useful
discussions. This work was partially supported by NSF Grant No.
DMR-0132874, NSF Career Program, MIT-Harvard CUA and AFOSR.

\end{document}